\documentclass[aip, rsi, amsmath, amssymb, reprint]{revtex4-1}

\usepackage{graphicx}
\usepackage{dcolumn}
\usepackage{bm}
\usepackage{units}

\begin{document}

\title{Geometric Dependence of Strong Field Enhanced Ionization in D$_{2}$O}

\author{Gregory A. McCracken}
\email{mccragre@stanford.edu}
\affiliation{ Stanford PULSE Institute, SLAC National Accelerator Laboratory\\
2575 Sand Hill Road, Menlo Park, CA 94025}
\affiliation{Department of Applied Physics, Stanford University, Stanford, CA 94305}

\author{Andreas Kaldun}
\affiliation{ Stanford PULSE Institute, SLAC National Accelerator Laboratory\\
2575 Sand Hill Road, Menlo Park, CA 94025}
\affiliation{Department of Physics, Stanford University, Stanford, CA 94305}

\author{Chelsea Liekhus-Schmaltz}
\affiliation{ Stanford PULSE Institute, SLAC National Accelerator Laboratory\\
2575 Sand Hill Road, Menlo Park, CA 94025}
\affiliation{Department of Physics, Stanford University, Stanford, CA 94305}

\author{Philip H. Bucksbaum}
\affiliation{ Stanford PULSE Institute, SLAC National Accelerator Laboratory\\
2575 Sand Hill Road, Menlo Park, CA 94025}
\affiliation{Department of Physics, Stanford University, Stanford, CA 94305}
\affiliation{Department of Applied Physics, Stanford University, Stanford, CA 94305}

\begin{abstract}
We have studied strong-field enhanced dissociative ionization of D$_{2}$O in 40 fs, 800 nm laser pulses with  focused intensities of {$\unitfrac[<1-3\times10^{15}]{W}{cm^{2}}$} by resolving the charged fragment momenta with respect to the laser polarization. 
We observe dication dissociation into OD$^{+}$/D$^{+}$ dominates when the polarization is out of the plane of the molecule, whereas trication dissociation into O$^{+}$/D$^{+}$/D$^{+}$ is strongly dominant when the polarization is aligned along the D-D axis.
Dication dissociation into O/D$^{+}$/D$^{+}$, and O$^{+}$/D$_2$$^{+}$ is not seen, nor is there any significant fragmentation into multiple ions when the laser is polarized along the C$_{2v}$ symmetry axis of the molecule. Even below the saturation intensity for OD$^{+}$/D$^{+}$, the O$^{+}$/D$^{+}$/D$^{+}$ channel has higher yield. 
By analyzing how the laser field is oriented within the molecular frame for both channels, we show that enhanced ionization is driving the triply charged three body breakup, but is not active for the doubly charged two body breakup. We conclude that laser-induced distortion of the molecular potential suppresses multiple ionization along the C$_{2v}$ axis, but enhances ionization along the D-D direction.

\end{abstract}

\date{\today}
\maketitle

\section{Introduction}
The ionization of molecules in strong ultrafast laser fields is a fundamental subject in laser-matter interactions, and also has relevance for laser chemistry and laser-induced  plasmas.\cite{seideman_role_1995,constant_observation_1996,gong_strong-field_2014,gong_strong-field_2014-1,Huang_2014,Lai_2016,Marangos_2016}, 
Strong-field molecular ionization is affected by vibrations and rotations: Laser-induced bond stretching tends to enhance ionization,\cite{zuo_charge-resonance-enhanced_1995}  while laser-driven bond alignment changes the polarization angle-dependent laser-electron coupling.\cite{stapelfeldt_colloquium_2003,tong_post_2005} Because of the strong-field coupling to these additional degrees of freedom, molecular ionization cannot be understood from simple low frequency tunneling theories developed for atomic systems.\cite{ammosov_tunnelling_1986} Modifications to tunneling ionization appropriate to molecular orbitals have been proposed,\cite{tong_theory_2002} but these become complicated in lighter molecules, where strong-field interactions with bonding orbitals can initiate fast nuclear motion leading to new phenomena such as induced multiple ionization.\cite{seideman_role_1995} 

To date, almost all experiments investigating the interplay between nuclear motion and ionization have involved linear molecules or symmetric tops\cite{Christiansen_2016,Cornaggia_2016}.
An exception is water, where strong-field induced electron removal from the inner-valence HOMO-1 bonding orbital initiates bending motion on the timescale of $\unit[800]{nm}$ laser field oscillations.  Studies comparing D$_2$O and H$_2$O have shown that this affects the spectrum in high harmonic generation (HHG).\cite{farrell_strong_2011} 

\begin{figure}[hb]
	\includegraphics[width = \columnwidth]{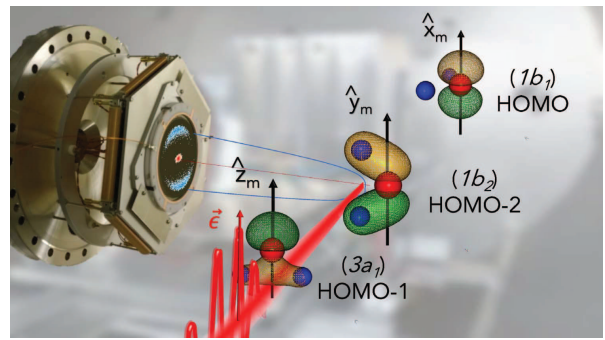}
    \caption{The three valence orbitals of D$_2$O are shown with the laser polarization aligned along the direction that has the highest probability of strong field ionization.  The dominant channel is found to be O+/D+/D+ at each intensity studied in this paper. Hits from this channel are shown on the detector. The molecular frame analysis suggests that the O+/D+/D+ channel is driven by alignment with the D-D bond, indicating the importance of the HOMO-2 orbital.}
	\label{fig:ExpCartoon}
\end{figure}

In this paper we have studied the dissociative multiple ionization of water in strong $\unit[800]{nm}$ focused laser fields. We examine dissociation channels where two or three electrons have been removed, and reconstruct the full momentum of all dissociating ions in coincidence, as well as the kinetic energy released in the coulomb explosion. Our results show how nuclear bending motion in the water molecule leads to dramatically enhanced multiple ionization in particular geometries, validating the general predictions of strong-field enhanced ionization.\cite{seideman_role_1995} The enhancement is  strongest when the laser is polarized along the D-D axis, where it couples most strongly to electrons in the HOMO-2 valence orbital (see Fig. \ref{fig:ExpCartoon}).  At the same time, we find that when the laser is polarized along the C$_{2v}$ axis where it couples most strongly to the HOMO-1 orbital, multiple ionization is nearly completely suppressed. While this second finding seems to contradict the general principles of enhanced ionization in strong fields, it may be due to a laser-induced shift in orbital energy and geometry as the molecule unbends.

\section{Experimental Methods}
The experiments are performed in a vacuum chamber at a base pressure of $\unit[3\times10^{-10}]{mbar}$ fitted with electrostatic focusing elements and a Roentdek time- and position-sensitive hex-anode delay-line detector.\cite{Jagutzki_2002}
The target is D$_{2}$O gas at 300K in the tight focus of a $\unit[40]{fs}$, $\unit[800]{nm}$ laser pulse. 
The focal volume and gas density yield fewer than one ionized molecule per laser pulse on average.  The electrostatic lenses direct all charged fragments to the detector, and the arrival time and position of each ion determines its mass and momentum. 

The intensity in the laser focus is scanned between $\unitfrac[7\times10^{14}]{W}{cm^{2}}$ and $\unitfrac[2\times10^{15}]{W}{cm^{2}}$, calibrated by comparing charge state ratios of argon in the same focus.\cite{bryan_atomic_2006} The laser propagation axis and polarization are in the detector plane, as shown in Fig.\ref{fig:ExpCartoon}. Data used in this analysis were accumulated over 250 million laser pulses at a $\unit[1]{kHz}$ repetition rate (69 hours). All ion hits were separately recorded for later analysis.

\section{Results and Analysis}
\begin{figure}
	\includegraphics[width = \columnwidth]{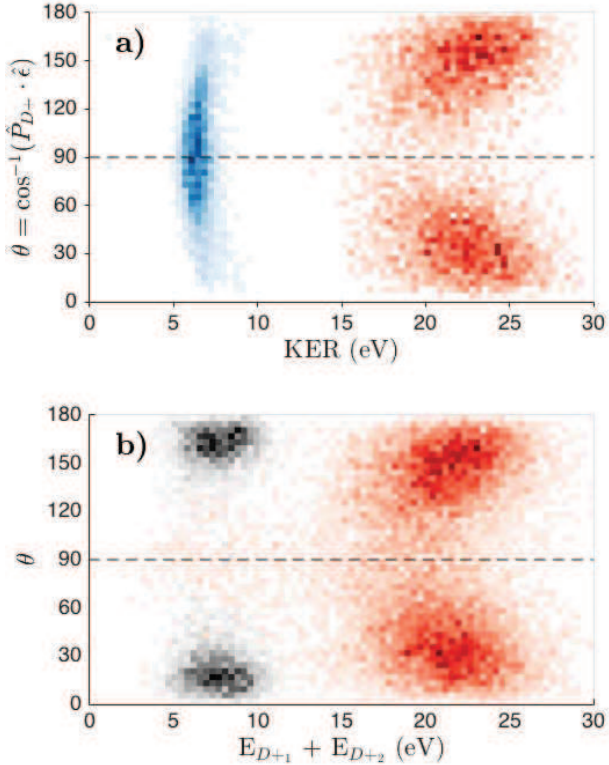}
	\caption{\textbf{a)} True coincidences from the OD$^{+}$/D$^{+}$ (blue) and O$^{+}$/D$^{+}$/D$^{+}$ (red) channels at an intensity of 2.1I$_{0}$, where I$_{0}$ = $\unitfrac[7\times10^{14}]{W}{cm^{2}}$. The kinetic energy release (KER) is plotted against $\theta$, the angle between the laser field and the trajectory of a randomly selected D$^{+}$ particle from the event. True coincidences are sorted for via momentum sums. \textbf{b)} $\theta$ versus the kinetic energy sum of D$^{+}$ fragments is plotted for all D$^{+}$/D$^{+}$ coincidences at 2.1I$_{0}$. Grey indicates events where the momentum sum goes to zero, and the KER is less than $\unit[12]{eV}$. These events are attributed to coulomb explosion of D$_2$, which is present in the D$_2$O sample. All other events are in red. The main contribution is from O$^{+}$/D$^{+}$/D$^{+}$, as seen by comparison with Fig.\ref{fig:Dall}a. There are no clear signatures of O/D$^{+}$/D$^{+}$ channels at lower KER.}
    \label{fig:Dall}
\end{figure}

The data were analyzed to identify all ionization channels that produce 2$^{+}$ or 3$^{+}$ charge states of D$_2$O. False coincidences, in which ions from more than one molecule were detected in the same shot, were efficiently eliminated by requiring that the momenta sum to nearly zero. This also misses partially ionized channels, including three-particle breakup where one of the particles was neutral such as O/D$^{+}$/D$^{+}$; but those can be recovered in a different way, described below. 
Of the four possible channels with non-neutral fragments, only two are seen:  OD$^{+}$/D$^{+}$ and O$^{+}$/D$^{+}$/D$^{+}$. Both channels can be seen easily in Fig.~\ref{fig:Dall}a, which plots the kinetic energy release (KER) against the angle $\theta$ between the momentum vector of one of the D$^{+}$ particles and the laser polarization. These two channels occupy distinct regions in this plot. 

The two-body dication decay O$^{+}$/D$_{2}^{+}$ is not seen, although it been noted in previous excitation studies: O$^{+}$/H$_{2}^{+}$ is a significant decay channel in $1s$ core exctation of water,\cite{laksman_rapid_2013} and also occurs in VUV excitation of the water cation.\cite{pedersen_photolysis_2013} This channel has also been reported in a previous strong-field ionization study, but the branching fraction was small, and could only be seen for the shortest pulses.\cite{rajgara_strong_2009}

We also looked for the dication breakup channel O/D$^{+}$/D$^{+}$. 
Although the two-body decay OD$^{+}$/D$^{+}$ is the most energetically favorable dissociation for double ionization, the three-body O/D$^{+}$/D$^{+}$ channel is known to be a significant dissociation pathway from some excited states of the dication.\cite{pedersen_photolysis_2013,streeter_classical_2017} 
For example, studies of single-photon double ionization of HOD at 40 eV observe O/H$^{+}$/D$^{+}$ breakup with a branching ratio of approximately 20\%.\cite{richardson_spectrum_1986}.  
If both of the ionized electrons are removed from inner valence orbitals via a strong-field process, then we might expect to see the O/D$^{+}$/D$^{+}$ channel.

Such partially ionized channels may be identified by their KER and orientation dependence, since the neutral particles are not detected.  To look for this, we plot all D$^{+}$/D$^{+}$ coincidences in Fig.\ref{fig:Dall}b. To filter out false coincidences with H$_{2}^{+}$, only events where each particle has more than $\unit[300]{meV}$ of kinetic energy are shown. Additionally, coincidences where the momentum sum of the D$^{+}$ particles is zero are shown in grey, and all other events are shown in red. One prominent feature are coincidences from O$^{+}$/D$^{+}$/D$^{+}$, which can be seen in red at high KER. There is one other channel at lower KER in grey, indicating the momentum sum is zero for these events. The distribution is practically identical to that of the H$^{+}$/H$^{+}$ channel measured in the same experiment, which can only be from the coulomb explosion of H$_{2}^{+}$. Therefore, the channel in grey is from the coulomb explosion of D$_{2}^{+}$, which is degassed from the heavy water sample. There are no other features in Fig.\ref{fig:Dall}b indicating the presence of a O/D$^{+}$/D$^{+}$ channel at this particular intensity, or for the range of intensities studied in this experiment. 

The dominance of the OD$^{+}$/D$^{+}$ channel suggests ionization to either the ground state or the first excited state in the dication, since these are the states that dissociate predominantly to OD$^{+}$/D$^{+}$.\cite{gervais_h2o2+_2009} These two states have electronic configurations with HOMO vacancies: $...(1b_{2})^{2}(3a_{1})^{1}(1b_{1})^{1}$ and $...(1b_{2})^{2}(3a_{1})^{2}(1b_{1})^{0}$ respectively. Therefore we expect these fragmentation products following strong field ionization of electrons from the HOMO.

The angle-dependent coupling of the molecular orbitals to the laser in the low-frequency limit is proportional to 
$F_{0}\hat{\epsilon}\cdot\vec{\mu}$, where $F_{0}\hat{\epsilon}$ is the laser field vector and $\vec{\mu}$ is the dipole moment of the orbital.\cite{tong_theory_2002}   For the $p$-type valence orbitals of water, the dipole moments are perpendicular to the nodal plane of the orbital.  As shown in Fig.\ref{fig:ExpCartoon}, this means that the ($1b_1$) HOMO orbital couples to the laser polarization along $\hat{x}_M$, 
perpendicular to the molecular plane; the ($3a_1$) HOMO-1 orbital couples to the laser polarization along $\hat{z}_M$, the C$_{2v}$ axis; and the ($1b_2$) HOMO-2 orbital couples to the laser polarization along $\hat{y}_M$, the D-D axis.  These three perpendicular axes $\hat{x}_M$, $\hat{y}_M$, and $\hat{z}_M$ in the molecular frame define the molecular alignment.

\begin{figure}[ht]
	\includegraphics[width = \columnwidth]{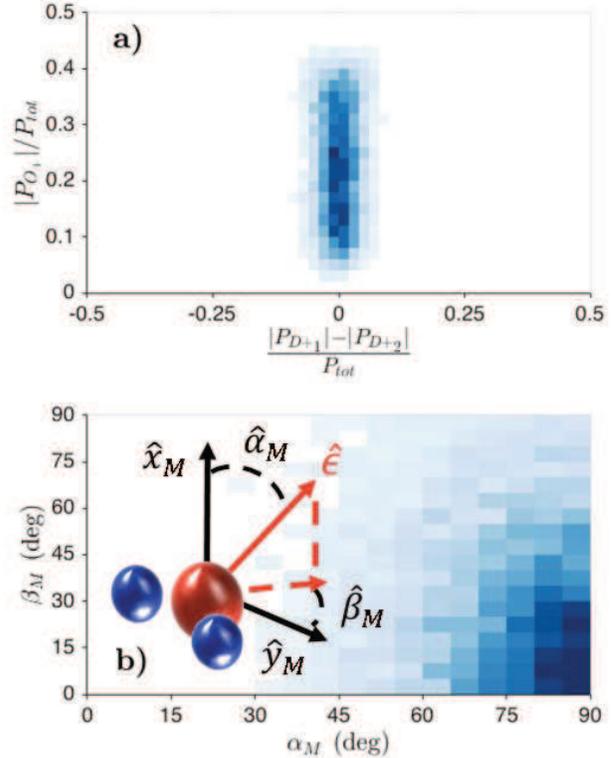}
	\caption{\textbf{a)} The O$^{+}$ momentum norm is plotted against the difference between D$^{+}$ momentum norms for the O$^{+}$/D$^{+}$/D$^{+}$ channel at an intensity of 2.1$I_0$. Both axes are normalized to sum of all momentum vector norms, P$_{tot}$. The vertical line centered at zero indicates a symmetric breakup. As the relative O$^{+}$ momentum approaches zero, the molecule dissociates in an increasingly linear geometry. \textbf{b)} Bivariate histogram of strong field orientation in the molecular frame for each O$^{+}$/D$^{+}$/D$^{+}$ event at 2.1I$_0$. This channel has a strong preference for alignment along the D-D bond. Orientation of the laser field with respect to the molecule is described by a polar angle, $\alpha_M$, and an azimuthal angle $\beta_M$.}
	\label{fig:Symmetric}
\end{figure}

For the OD$^{+}$/D$^{+}$ channel, Fig.\ref{fig:Dall}a shows that the dissociation axis is predominantly perpendicular to the laser polarization, i.e. the strong field of the laser is orthogonal to the molecular plane. Since the polarizability of the lone-pair HOMO is along this direction, ionization from this orbital likely plays a role (see Fig.\ref{fig:ExpCartoon}). Ionization from the HOMO is involved in the excitation of the lowest dication states.\cite{van_huis_scratching_1999,nobusada_theoretical_2000} Furthermore, the KER of the OD$^{+}$/D$^{+}$ channel is similar to that found in PIPICO studies using single photon excitation near the vertical excitation energies of the lowest dication states.\cite{richardson_spectrum_1986,winkoun_one-_1988} The KER also has a relatively narrow distribution. This indicates that there is minimal nuclear motion during the ionization. Indeed, removal of a HOMO electron prepares the cation in the ground state which has a very similar equilibrium geometry to the neutral ground state.\cite{schneider_lower_1996,brundle_high_1968} Therefore, alignment of the laser field out of the molecular plane drives ionization of the HOMO, and leads to OD$^{+}$/D$^{+}$ dissociation.

While OD$^{+}$/D$^{+}$ dissociation is driven by alignment of the strong field normal to the molecular plane, O$^{+}$/D$^{+}$/D$^{+}$ dissociation is driven by alignment of the field in the molecular plane. To show dependence on alignment, we reconstruct the molecular frame for each event by taking advantage of the symmetry of the dissociation. The orientation with respect to the laser field can then be determined.
Fig.\ref{fig:Symmetric}a shows the normalized O$^{+}$ momentum versus the difference of normalized D$^{+}$ momentum for each event in the channel. The vertical line at a D$^{+}$ momentum difference of zero indicates that the breakup is symmetric. Because of the symmetry, we can use the momenta of the fragments to reconstruct the molecular axes in a straightforward manner.

The molecular axes in the lab frame are then determined by the O$^{+}$ momentum ($\hat{z}_M$), and the cross product of D$^{+}$ momenta ($\hat{x}_M$). $\hat{y}_M$ is found by completing the right handed coordinate system. For each event, the laser polarization is oriented within the molecular frame. Fig.\ref{fig:Symmetric}b shows the results, along with a description of the spherical coordinate system used to describe the orientation of the laser field. The analysis shows that the polarization is primarily in the molecular plane for the O$^{+}$/D$^{+}$/D$^{+}$ channel, and preferentially along the D-D bond. Therefore, in plane alignment of the strong field leads to triple ionization and a symmetric dissociation, and out of plane alignment leads to double ionization and a asymmetric fragmentation into OD$^{+}$/D$^{+}$.

\begin{figure}[h]
	\includegraphics[width = \columnwidth]{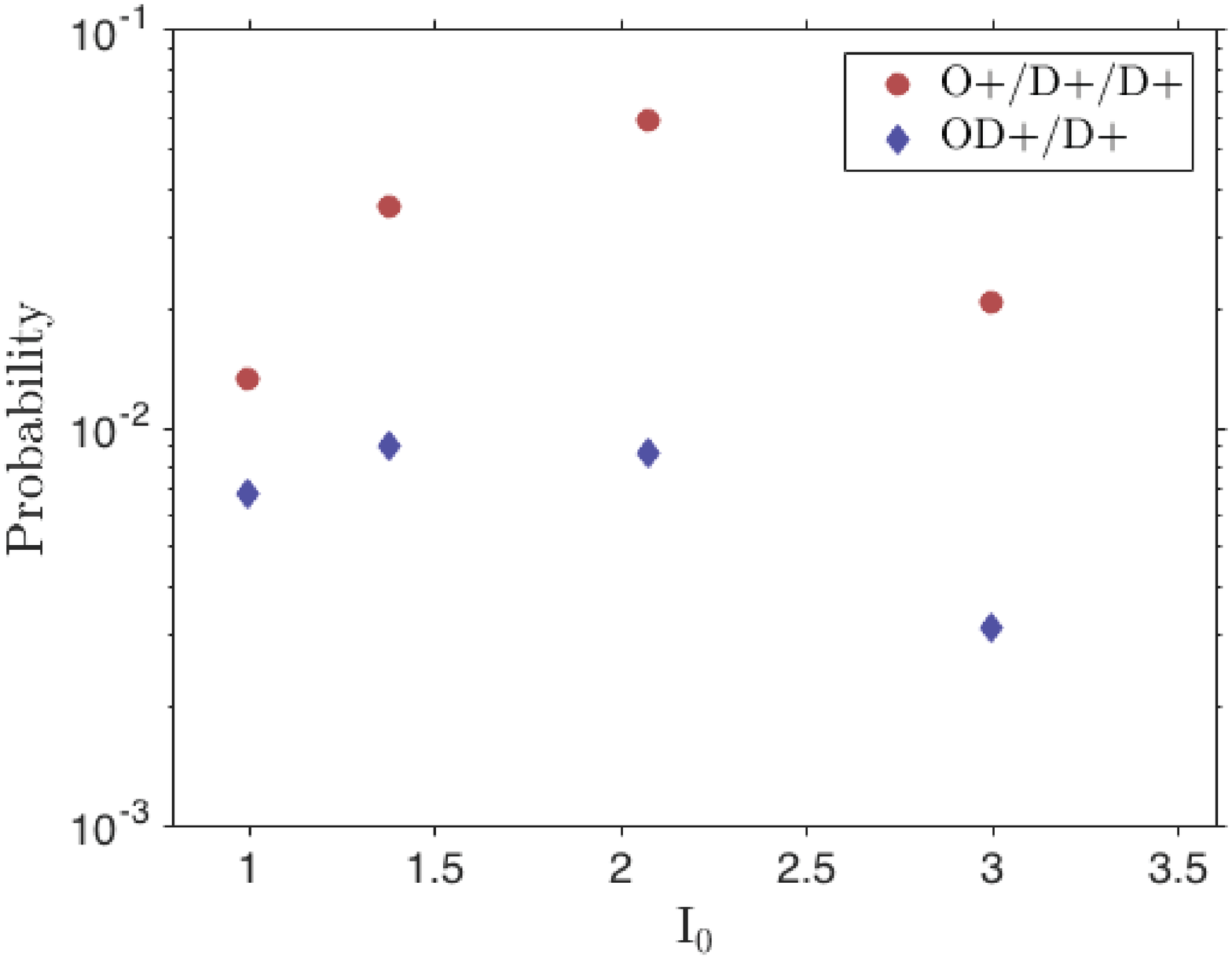}
	\caption{Probability for OD$^{+}$/D$^{+}$ and O$^{+}$/D$^{+}$/D$^{+}$ in D$_2$O versus intensity. Since OD$^{+}$/D$^{+}$ and O$^{+}$/D$^{+}$/D$^{+}$ are the only dissociation channels that were detected, these are the probabilities for double and triple ionization. Surprisingly, even below saturation for double ionization, triple ionization is always more likely. Intensity is in units of I$_0$ = $\unitfrac[7\times10^{14}]{W}{cm^{2}}$.}
	\label{fig:prob}
\end{figure}

\begin{figure}[hb]
	\includegraphics[width = \columnwidth]{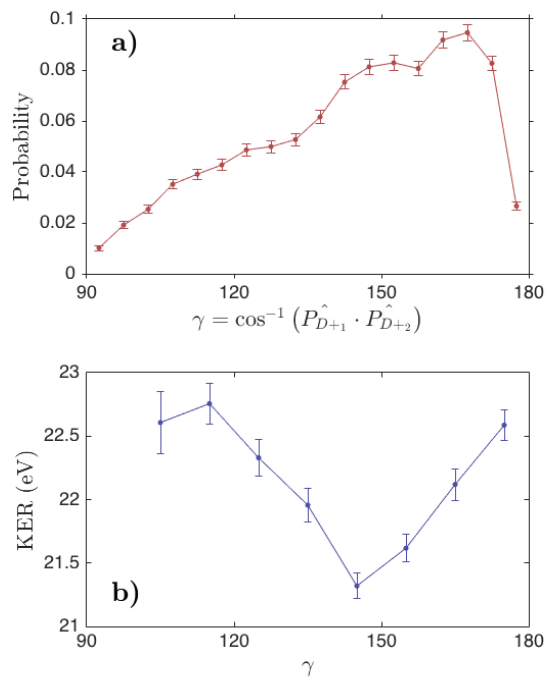}
	\caption{\textbf{a)} The dissociation angle, $\gamma$, measured in the O$^{+}$/D$^{+}$/D$^{+}$ channel for 2.1I$_{0}$, related to the bond angle upon dissociation. There is a substructure indicating multiple dissociation pathways, with most pathways involving increase in the bond angle. \textbf{b)} The mean KER for different dissociation angles obtained by Gaussian fits to the KER spectrum for 10 degree bins. The non-monotonic scaling with angle further indicates there are different subchannels.}
    \label{fig:MultiChan}
\end{figure}

We have seen that the field orientation plays a large role in the yield for both channels. More strikingly, double ionization is significantly less probable than triple ionization over the whole range of intensities used here. The probabilities of the OD$^{+}$/D$^{+}$ and O$^{+}$/D$^{+}$/D$^{+}$ channels across different intensities are shown in Fig.~\ref{fig:prob}. Since these are the only two channels with appreciable yield for each intensity, and there are no stable states of the dication,\cite{van_huis_scratching_1999} the yield of each channel gives the probability for double and triple ionization. Triple ionization is always more probable than double ionization, even before the double ionization saturation point between 1.5I$_0$ and 2.0I$_0$. This is a dramatic departure from what is expected in low frequency, strong field ionization. Below saturation for removing N-1 electrons, ionization of N-1 electrons should be more likely than ionization of N electrons. This is not the case for D$_2$O.

The unexpectedly high yield for the O$^{+}$/D$^{+}$/D$^{+}$ channel could be explained by enhanced ionization. Enhanced ionization is driven by the laser field's interaction with the polarizability of the molecule. The interaction, which is strongest when the field is aligned with the polarizability, must occur along nuclear degrees of freedom. For the OD$^{+}$/D$^{+}$ channel, the laser is predominantly oriented along $\hat{x}_{M}$, so cannot effectively induce nuclear motion. Conversely, O$^{+}$/D$^{+}$/D$^{+}$ is driven by in plane alignment of the field. Therefore, the OD$^{+}$/D$^{+}$ channel does not involve enhanced ionization, while the effect is likely active for O$^{+}$/D$^{+}$/D$^{+}$. Evidence of enhanced ionization has been seen before in water,~\cite{sanderson_geometry_1999,legare_laser_2005} where stretching of the angular bond was shown to be important. Fig.~\ref{fig:MultiChan}a shows the distribution of dissociation angles, $\gamma$, for the O$^{+}$/D$^{+}$/D$^{+}$ channel. This dissociation angle is related to the bond angle upon coulomb explosion. Significant angular stretching can be seen to occur, especially at higher intensity. Since the laser field is predominantly aligned to the D-D bond in this channel, it can interact with the polarizability of the HOMO-2 and cause the molecule to become more linear.

Sub-structure can be seen in the $\gamma$ distribution, indicating multiple pathways to enhanced ionization. Fig.~\ref{fig:MultiChan}b shows that the KER varies with $\gamma$ in a non-monotonic manner. This further suggests that there are different pathways to enhanced ionization with unique critical values for both angular and O-D bonds.

Although enhanced ionization may explain the high yield of O$^{+}$/D$^{+}$/D$^{+}$ relative to OD$^{+}$/D$^{+}$, it does not explain why O/D$^{+}$/D$^{+}$ does not have appreciable probability. Across all intensities, there is scare evidence of O/D$^{+}$/D$^{+}$, much like in Fig.\ref{fig:Dall}b. This channel is known to be a significant pathway for dication states that are excited by the removal of electrons from lower orbitals.\cite{cooper_study_1987,streeter_classical_2017} Since the laser field is in the molecular plane for enhanced ionization, ionization involves removal of the electrons from inner orbitals of D$_2$O, so the dications created are in excited states. 
Below the saturation point for triple ionization, which is around 2.1I$_0$ according to Fig.~\ref{fig:prob}, there should be appreciable probability for O/D$^{+}$/D$^{+}$. Since this is not seen, removal of a third electron may be driven by autoionization, or multiphoton resonant ionization pathways.

Enhanced ionization is also suppressed for field alignment along $\hat{z}_M$, as shown in Fig.\ref{fig:Symmetric}b. This is peculiar since field alignment along $\hat{z}_M$ would allow for stretching of the OD bonds due to interaction with the HOMO-1. Studies of isoelectronic H$_2$S in high intensity $\unit[800]{nm}$ pulses have shown that ionization rates are also significantly lower with the laser polarization along the symmetry axis.\cite{hishikawa_electronic_2006} Effects such as the re-ordering of orbital energies due to the strong field driven nuclear distortion should be considered in understanding the suppression of enhanced ionization in bent triatomics along a particular in-plane axis. 

\section{Conclusion}

In this work, we saw two dominant ionization pathways for strong-field double and triple ionization from D$_{2}$O: an asymmetric OD$^{+}$/D$^{+}$ channel driven by laser field alignment orthogonal to the molecular plane, and a symmetric O$^{+}$/D$^{+}$/D$^{+}$ channel driven by field alignment along the D-D bond. Triple ionization was found to be more likely than double ionization, even below the saturation intensity for double ionization. This is partly due to enhanced ionization only being active for the O$^{+}$/D$^{+}$/D$^{+}$ channel, where the laser field can effectively induce bond stretching. However, it is curious why the O/D$^{+}$/D$^{+}$ channel cannot be a product of enhanced ionization as it is a primary dissociation pathway in the dication. Mechanisms such as autoionization or multiphoton resonant ionization might be involved in the ionization process when the strong field is in the molecular plane. Furthermore, the apparent suppression of enhanced ionization along the symmetry axis is intriguing. It is worth investigating if suppression of enhanced ionization along a particular axis is a general feature in bent triatomics.

This work was supported by the National Science Foundation under Grant No. PHY-0649578.

\bibliography{My_Library}{}

\begin{thebibliography}{32}%
\makeatletter
\providecommand \@ifxundefined [1]{%
 \@ifx{#1\undefined}
}%
\providecommand \@ifnum [1]{%
 \ifnum #1\expandafter \@firstoftwo
 \else \expandafter \@secondoftwo
 \fi
}%
\providecommand \@ifx [1]{%
 \ifx #1\expandafter \@firstoftwo
 \else \expandafter \@secondoftwo
 \fi
}%
\providecommand \natexlab [1]{#1}%
\providecommand \enquote  [1]{``#1''}%
\providecommand \bibnamefont  [1]{#1}%
\providecommand \bibfnamefont [1]{#1}%
\providecommand \citenamefont [1]{#1}%
\providecommand \href@noop [0]{\@secondoftwo}%
\providecommand \href [0]{\begingroup \@sanitize@url \@href}%
\providecommand \@href[1]{\@@startlink{#1}\@@href}%
\providecommand \@@href[1]{\endgroup#1\@@endlink}%
\providecommand \@sanitize@url [0]{\catcode `\\12\catcode `\$12\catcode
  `\&12\catcode `\#12\catcode `\^12\catcode `\_12\catcode `\%12\relax}%
\providecommand \@@startlink[1]{}%
\providecommand \@@endlink[0]{}%
\providecommand \url  [0]{\begingroup\@sanitize@url \@url }%
\providecommand \@url [1]{\endgroup\@href {#1}{\urlprefix }}%
\providecommand \urlprefix  [0]{URL }%
\providecommand \Eprint [0]{\href }%
\providecommand \doibase [0]{http://dx.doi.org/}%
\providecommand \selectlanguage [0]{\@gobble}%
\providecommand \bibinfo  [0]{\@secondoftwo}%
\providecommand \bibfield  [0]{\@secondoftwo}%
\providecommand \translation [1]{[#1]}%
\providecommand \BibitemOpen [0]{}%
\providecommand \bibitemStop [0]{}%
\providecommand \bibitemNoStop [0]{.\EOS\space}%
\providecommand \EOS [0]{\spacefactor3000\relax}%
\providecommand \BibitemShut  [1]{\csname bibitem#1\endcsname}%
\let\auto@bib@innerbib\@empty
\bibitem [{\citenamefont {Seideman}, \citenamefont {Ivanov},\ and\
  \citenamefont {Corkum}(1995)}]{seideman_role_1995}%
  \BibitemOpen
  \bibfield  {author} {\bibinfo {author} {\bibfnamefont {T.}~\bibnamefont
  {Seideman}}, \bibinfo {author} {\bibfnamefont {M.~Y.}\ \bibnamefont
  {Ivanov}}, \ and\ \bibinfo {author} {\bibfnamefont {P.~B.}\ \bibnamefont
  {Corkum}},\ }\href {\doibase 10.1103/PhysRevLett.75.2819} {\bibfield
  {journal} {\bibinfo  {journal} {Physical Review Letters}\ }\textbf {\bibinfo
  {volume} {75}},\ \bibinfo {pages} {2819} (\bibinfo {year}
  {1995})}\BibitemShut {NoStop}%
\bibitem [{\citenamefont {Constant}, \citenamefont {Stapelfeldt},\ and\
  \citenamefont {Corkum}(1996)}]{constant_observation_1996}%
  \BibitemOpen
  \bibfield  {author} {\bibinfo {author} {\bibfnamefont {E.}~\bibnamefont
  {Constant}}, \bibinfo {author} {\bibfnamefont {H.}~\bibnamefont
  {Stapelfeldt}}, \ and\ \bibinfo {author} {\bibfnamefont {P.~B.}\ \bibnamefont
  {Corkum}},\ }\href {\doibase 10.1103/PhysRevLett.76.4140} {\bibfield
  {journal} {\bibinfo  {journal} {Physical Review Letters}\ }\textbf {\bibinfo
  {volume} {76}},\ \bibinfo {pages} {4140} (\bibinfo {year}
  {1996})}\BibitemShut {NoStop}%
\bibitem [{\citenamefont {Gong}\ \emph
  {et~al.}(2014{\natexlab{a}})\citenamefont {Gong}, \citenamefont {Song},
  \citenamefont {Ji}, \citenamefont {Pan}, \citenamefont {Ding}, \citenamefont
  {Wu},\ and\ \citenamefont {Zeng}}]{gong_strong-field_2014}%
  \BibitemOpen
  \bibfield  {author} {\bibinfo {author} {\bibfnamefont {X.}~\bibnamefont
  {Gong}}, \bibinfo {author} {\bibfnamefont {Q.}~\bibnamefont {Song}}, \bibinfo
  {author} {\bibfnamefont {Q.}~\bibnamefont {Ji}}, \bibinfo {author}
  {\bibfnamefont {H.}~\bibnamefont {Pan}}, \bibinfo {author} {\bibfnamefont
  {J.}~\bibnamefont {Ding}}, \bibinfo {author} {\bibfnamefont {J.}~\bibnamefont
  {Wu}}, \ and\ \bibinfo {author} {\bibfnamefont {H.}~\bibnamefont {Zeng}},\
  }\href {\doibase 10.1103/PhysRevLett.112.243001} {\bibfield  {journal}
  {\bibinfo  {journal} {Physical Review Letters}\ }\textbf {\bibinfo {volume}
  {112}} (\bibinfo {year} {2014}{\natexlab{a}}),\
  10.1103/PhysRevLett.112.243001}\BibitemShut {NoStop}%
\bibitem [{\citenamefont {Gong}\ \emph
  {et~al.}(2014{\natexlab{b}})\citenamefont {Gong}, \citenamefont {Song},
  \citenamefont {Ji}, \citenamefont {Pan}, \citenamefont {Ding}, \citenamefont
  {Wu},\ and\ \citenamefont {Zeng}}]{gong_strong-field_2014-1}%
  \BibitemOpen
  \bibfield  {author} {\bibinfo {author} {\bibfnamefont {X.}~\bibnamefont
  {Gong}}, \bibinfo {author} {\bibfnamefont {Q.}~\bibnamefont {Song}}, \bibinfo
  {author} {\bibfnamefont {Q.}~\bibnamefont {Ji}}, \bibinfo {author}
  {\bibfnamefont {H.}~\bibnamefont {Pan}}, \bibinfo {author} {\bibfnamefont
  {J.}~\bibnamefont {Ding}}, \bibinfo {author} {\bibfnamefont {J.}~\bibnamefont
  {Wu}}, \ and\ \bibinfo {author} {\bibfnamefont {H.}~\bibnamefont {Zeng}},\
  }\href {\doibase 10.1103/PhysRevLett.112.243001} {\bibfield  {journal}
  {\bibinfo  {journal} {Physical Review Letters}\ }\textbf {\bibinfo {volume}
  {112}} (\bibinfo {year} {2014}{\natexlab{b}}),\
  10.1103/PhysRevLett.112.243001}\BibitemShut {NoStop}%
\bibitem [{\citenamefont {Huang}(2014)}]{Huang_2014}%
  \BibitemOpen
  \bibfield  {author} {\bibinfo {author} {\bibfnamefont {C.}~\bibnamefont
  {Huang}},\ }\href {\doibase 10.1103/PhysRevA.90.043420} {\bibfield  {journal}
  {\bibinfo  {journal} {Physical Review A}\ }\textbf {\bibinfo {volume} {90}}
  (\bibinfo {year} {2014}),\ 10.1103/PhysRevA.90.043420}\BibitemShut {NoStop}%
\bibitem [{\citenamefont {Lai}(2016)}]{Lai_2016}%
  \BibitemOpen
  \bibfield  {author} {\bibinfo {author} {\bibfnamefont {W.}~\bibnamefont
  {Lai}},\ }\href {\doibase 10.1103/PhysRevA.93.043401} {\bibfield  {journal}
  {\bibinfo  {journal} {Physical Review A}\ }\textbf {\bibinfo {volume} {93}}
  (\bibinfo {year} {2016}),\ 10.1103/PhysRevA.93.043401}\BibitemShut {NoStop}%
\bibitem [{\citenamefont {Schütte}\ \emph {et~al.}(2016)\citenamefont
  {Schütte}, \citenamefont {Ye}, \citenamefont {Patchkovskii}, \citenamefont
  {Austin}, \citenamefont {Brahms}, \citenamefont {Strüber}, \citenamefont
  {Witting}, \citenamefont {Ivanov}, \citenamefont {Tisch},\ and\ \citenamefont
  {Marangos}}]{Marangos_2016}%
  \BibitemOpen
  \bibfield  {author} {\bibinfo {author} {\bibfnamefont {B.}~\bibnamefont
  {Schütte}}, \bibinfo {author} {\bibfnamefont {P.}~\bibnamefont {Ye}},
  \bibinfo {author} {\bibfnamefont {S.}~\bibnamefont {Patchkovskii}}, \bibinfo
  {author} {\bibfnamefont {D.~R.}\ \bibnamefont {Austin}}, \bibinfo {author}
  {\bibfnamefont {C.}~\bibnamefont {Brahms}}, \bibinfo {author} {\bibfnamefont
  {C.}~\bibnamefont {Strüber}}, \bibinfo {author} {\bibfnamefont
  {T.}~\bibnamefont {Witting}}, \bibinfo {author} {\bibfnamefont {M.~Y.}\
  \bibnamefont {Ivanov}}, \bibinfo {author} {\bibfnamefont {J.~W.~G.}\
  \bibnamefont {Tisch}}, \ and\ \bibinfo {author} {\bibfnamefont {J.~P.}\
  \bibnamefont {Marangos}},\ }\href@noop {} {\bibfield  {journal} {\bibinfo
  {journal} {Scientific Reports}\ }\textbf {\bibinfo {volume} {6}},\ \bibinfo
  {pages} {srep39664} (\bibinfo {year} {2016})}\BibitemShut {NoStop}%
\bibitem [{\citenamefont {Zuo}\ and\ \citenamefont
  {Bandrauk}(1995)}]{zuo_charge-resonance-enhanced_1995}%
  \BibitemOpen
  \bibfield  {author} {\bibinfo {author} {\bibfnamefont {T.}~\bibnamefont
  {Zuo}}\ and\ \bibinfo {author} {\bibfnamefont {A.~D.}\ \bibnamefont
  {Bandrauk}},\ }\href {\doibase 10.1103/PhysRevA.52.R2511} {\bibfield
  {journal} {\bibinfo  {journal} {Physical Review A}\ }\textbf {\bibinfo
  {volume} {52}},\ \bibinfo {pages} {R2511} (\bibinfo {year}
  {1995})}\BibitemShut {NoStop}%
\bibitem [{\citenamefont {Stapelfeldt}\ and\ \citenamefont
  {Seideman}(2003)}]{stapelfeldt_colloquium_2003}%
  \BibitemOpen
  \bibfield  {author} {\bibinfo {author} {\bibfnamefont {H.}~\bibnamefont
  {Stapelfeldt}}\ and\ \bibinfo {author} {\bibfnamefont {T.}~\bibnamefont
  {Seideman}},\ }\href {\doibase 10.1103/RevModPhys.75.543} {\bibfield
  {journal} {\bibinfo  {journal} {Reviews of Modern Physics}\ }\textbf
  {\bibinfo {volume} {75}},\ \bibinfo {pages} {543} (\bibinfo {year}
  {2003})}\BibitemShut {NoStop}%
\bibitem [{\citenamefont {Tong}\ \emph {et~al.}(2005)\citenamefont {Tong},
  \citenamefont {Zhao}, \citenamefont {Alnaser}, \citenamefont {Voss},
  \citenamefont {Cocke},\ and\ \citenamefont {Lin}}]{tong_post_2005}%
  \BibitemOpen
  \bibfield  {author} {\bibinfo {author} {\bibfnamefont {X.~M.}\ \bibnamefont
  {Tong}}, \bibinfo {author} {\bibfnamefont {Z.~X.}\ \bibnamefont {Zhao}},
  \bibinfo {author} {\bibfnamefont {A.~S.}\ \bibnamefont {Alnaser}}, \bibinfo
  {author} {\bibfnamefont {S.}~\bibnamefont {Voss}}, \bibinfo {author}
  {\bibfnamefont {C.~L.}\ \bibnamefont {Cocke}}, \ and\ \bibinfo {author}
  {\bibfnamefont {C.~D.}\ \bibnamefont {Lin}},\ }\href {\doibase
  10.1088/0953-4075/38/4/002} {\bibfield  {journal} {\bibinfo  {journal}
  {Journal of Physics B: Atomic, Molecular and Optical Physics}\ }\textbf
  {\bibinfo {volume} {38}},\ \bibinfo {pages} {333} (\bibinfo {year}
  {2005})}\BibitemShut {NoStop}%
\bibitem [{\citenamefont {Ammosov}, \citenamefont {Delone},\ and\ \citenamefont
  {Krainov}(1986)}]{ammosov_tunnelling_1986}%
  \BibitemOpen
  \bibfield  {author} {\bibinfo {author} {\bibfnamefont {M.}~\bibnamefont
  {Ammosov}}, \bibinfo {author} {\bibfnamefont {N.}~\bibnamefont {Delone}}, \
  and\ \bibinfo {author} {\bibfnamefont {V.}~\bibnamefont {Krainov}},\
  }\href@noop {} {\bibfield  {journal} {\bibinfo  {journal} {Sov. Phys. JETP}\
  }\textbf {\bibinfo {volume} {64}},\ \bibinfo {pages} {1191} (\bibinfo {year}
  {1986})}\BibitemShut {NoStop}%
\bibitem [{\citenamefont {Tong}, \citenamefont {Zhao},\ and\ \citenamefont
  {Lin}(2002)}]{tong_theory_2002}%
  \BibitemOpen
  \bibfield  {author} {\bibinfo {author} {\bibfnamefont {X.~M.}\ \bibnamefont
  {Tong}}, \bibinfo {author} {\bibfnamefont {Z.~X.}\ \bibnamefont {Zhao}}, \
  and\ \bibinfo {author} {\bibfnamefont {C.~D.}\ \bibnamefont {Lin}},\ }\href
  {\doibase 10.1103/PhysRevA.66.033402} {\bibfield  {journal} {\bibinfo
  {journal} {Physical Review A}\ }\textbf {\bibinfo {volume} {66}} (\bibinfo
  {year} {2002}),\ 10.1103/PhysRevA.66.033402}\BibitemShut {NoStop}%
\bibitem [{\citenamefont {Christiansen}(2016)}]{Christiansen_2016}%
  \BibitemOpen
  \bibfield  {author} {\bibinfo {author} {\bibfnamefont {L.}~\bibnamefont
  {Christiansen}},\ }\href {\doibase 10.1103/PhysRevA.93.023411} {\bibfield
  {journal} {\bibinfo  {journal} {Physical Review A}\ }\textbf {\bibinfo
  {volume} {93}} (\bibinfo {year} {2016}),\
  10.1103/PhysRevA.93.023411}\BibitemShut {NoStop}%
\bibitem [{\citenamefont {Cornaggia}(2016)}]{Cornaggia_2016}%
  \BibitemOpen
  \bibfield  {author} {\bibinfo {author} {\bibfnamefont {C.}~\bibnamefont
  {Cornaggia}},\ }\href {\doibase 10.1088/0953-4075/49/19/19LT01} {\bibfield
  {journal} {\bibinfo  {journal} {Journal of Physics B: Atomic, Molecular and
  Optical Physics}\ }\textbf {\bibinfo {volume} {49}},\ \bibinfo {pages}
  {19LT01} (\bibinfo {year} {2016})}\BibitemShut {NoStop}%
\bibitem [{\citenamefont {Farrell}\ \emph {et~al.}(2011)\citenamefont
  {Farrell}, \citenamefont {Petretti}, \citenamefont {Förster}, \citenamefont
  {McFarland}, \citenamefont {Spector}, \citenamefont {Vanne}, \citenamefont
  {Decleva}, \citenamefont {Bucksbaum}, \citenamefont {Saenz},\ and\
  \citenamefont {Gühr}}]{farrell_strong_2011}%
  \BibitemOpen
  \bibfield  {author} {\bibinfo {author} {\bibfnamefont {J.~P.}\ \bibnamefont
  {Farrell}}, \bibinfo {author} {\bibfnamefont {S.}~\bibnamefont {Petretti}},
  \bibinfo {author} {\bibfnamefont {J.}~\bibnamefont {Förster}}, \bibinfo
  {author} {\bibfnamefont {B.~K.}\ \bibnamefont {McFarland}}, \bibinfo {author}
  {\bibfnamefont {L.~S.}\ \bibnamefont {Spector}}, \bibinfo {author}
  {\bibfnamefont {Y.~V.}\ \bibnamefont {Vanne}}, \bibinfo {author}
  {\bibfnamefont {P.}~\bibnamefont {Decleva}}, \bibinfo {author} {\bibfnamefont
  {P.~H.}\ \bibnamefont {Bucksbaum}}, \bibinfo {author} {\bibfnamefont
  {A.}~\bibnamefont {Saenz}}, \ and\ \bibinfo {author} {\bibfnamefont
  {M.}~\bibnamefont {Gühr}},\ }\href {\doibase 10.1103/PhysRevLett.107.083001}
  {\bibfield  {journal} {\bibinfo  {journal} {Physical Review Letters}\
  }\textbf {\bibinfo {volume} {107}} (\bibinfo {year} {2011}),\
  10.1103/PhysRevLett.107.083001}\BibitemShut {NoStop}%
\bibitem [{\citenamefont {Jagutzki}\ \emph {et~al.}(2002)\citenamefont
  {Jagutzki}, \citenamefont {Cerezo}, \citenamefont {Czasch}, \citenamefont
  {Dorner}, \citenamefont {Hattas}, \citenamefont {Huang}, \citenamefont
  {Mergel}, \citenamefont {Spillmann}, \citenamefont {Ullmann-Pfleger},
  \citenamefont {Weber},\ and\ \citenamefont {et~al.}}]{Jagutzki_2002}%
  \BibitemOpen
  \bibfield  {author} {\bibinfo {author} {\bibfnamefont {O.}~\bibnamefont
  {Jagutzki}}, \bibinfo {author} {\bibfnamefont {A.}~\bibnamefont {Cerezo}},
  \bibinfo {author} {\bibfnamefont {A.}~\bibnamefont {Czasch}}, \bibinfo
  {author} {\bibfnamefont {R.}~\bibnamefont {Dorner}}, \bibinfo {author}
  {\bibfnamefont {M.}~\bibnamefont {Hattas}}, \bibinfo {author} {\bibfnamefont
  {M.}~\bibnamefont {Huang}}, \bibinfo {author} {\bibfnamefont
  {V.}~\bibnamefont {Mergel}}, \bibinfo {author} {\bibfnamefont
  {U.}~\bibnamefont {Spillmann}}, \bibinfo {author} {\bibfnamefont
  {K.}~\bibnamefont {Ullmann-Pfleger}}, \bibinfo {author} {\bibfnamefont
  {T.}~\bibnamefont {Weber}}, \ and\ \bibinfo {author} {\bibnamefont
  {et~al.}},\ }\href {\doibase 10.1109/TNS.2002.803889} {\bibfield  {journal}
  {\bibinfo  {journal} {IEEE Transactions on Nuclear Science}\ }\textbf
  {\bibinfo {volume} {49}},\ \bibinfo {pages} {2477–2483} (\bibinfo {year}
  {2002})}\BibitemShut {NoStop}%
\bibitem [{\citenamefont {Bryan}\ \emph {et~al.}(2006)\citenamefont {Bryan},
  \citenamefont {Stebbings}, \citenamefont {McKenna}, \citenamefont {English},
  \citenamefont {Suresh}, \citenamefont {Wood}, \citenamefont {Srigengan},
  \citenamefont {Turcu}, \citenamefont {Smith}, \citenamefont {Divall},
  \citenamefont {Hooker}, \citenamefont {Langley}, \citenamefont {Collier},
  \citenamefont {Williams},\ and\ \citenamefont {Newell}}]{bryan_atomic_2006}%
  \BibitemOpen
  \bibfield  {author} {\bibinfo {author} {\bibfnamefont {W.~A.}\ \bibnamefont
  {Bryan}}, \bibinfo {author} {\bibfnamefont {S.~L.}\ \bibnamefont
  {Stebbings}}, \bibinfo {author} {\bibfnamefont {J.}~\bibnamefont {McKenna}},
  \bibinfo {author} {\bibfnamefont {E.~M.~L.}\ \bibnamefont {English}},
  \bibinfo {author} {\bibfnamefont {M.}~\bibnamefont {Suresh}}, \bibinfo
  {author} {\bibfnamefont {J.}~\bibnamefont {Wood}}, \bibinfo {author}
  {\bibfnamefont {B.}~\bibnamefont {Srigengan}}, \bibinfo {author}
  {\bibfnamefont {I.~C.~E.}\ \bibnamefont {Turcu}}, \bibinfo {author}
  {\bibfnamefont {J.~M.}\ \bibnamefont {Smith}}, \bibinfo {author}
  {\bibfnamefont {E.~J.}\ \bibnamefont {Divall}}, \bibinfo {author}
  {\bibfnamefont {C.~J.}\ \bibnamefont {Hooker}}, \bibinfo {author}
  {\bibfnamefont {A.~J.}\ \bibnamefont {Langley}}, \bibinfo {author}
  {\bibfnamefont {J.~L.}\ \bibnamefont {Collier}}, \bibinfo {author}
  {\bibfnamefont {I.~D.}\ \bibnamefont {Williams}}, \ and\ \bibinfo {author}
  {\bibfnamefont {W.~R.}\ \bibnamefont {Newell}},\ }\href {\doibase
  10.1038/nphys310} {\bibfield  {journal} {\bibinfo  {journal} {Nature
  Physics}\ }\textbf {\bibinfo {volume} {2}},\ \bibinfo {pages} {379} (\bibinfo
  {year} {2006})}\BibitemShut {NoStop}%
\bibitem [{\citenamefont {Laksman}\ \emph {et~al.}(2013)\citenamefont
  {Laksman}, \citenamefont {Månsson}, \citenamefont {Sankari}, \citenamefont
  {Céolin}, \citenamefont {Gisselbrecht},\ and\ \citenamefont
  {Sorensen}}]{laksman_rapid_2013}%
  \BibitemOpen
  \bibfield  {author} {\bibinfo {author} {\bibfnamefont {J.}~\bibnamefont
  {Laksman}}, \bibinfo {author} {\bibfnamefont {E.~P.}\ \bibnamefont
  {Månsson}}, \bibinfo {author} {\bibfnamefont {A.}~\bibnamefont {Sankari}},
  \bibinfo {author} {\bibfnamefont {D.}~\bibnamefont {Céolin}}, \bibinfo
  {author} {\bibfnamefont {M.}~\bibnamefont {Gisselbrecht}}, \ and\ \bibinfo
  {author} {\bibfnamefont {S.~L.}\ \bibnamefont {Sorensen}},\ }\href {\doibase
  10.1039/c3cp52625a} {\bibfield  {journal} {\bibinfo  {journal} {Physical
  Chemistry Chemical Physics}\ }\textbf {\bibinfo {volume} {15}},\ \bibinfo
  {pages} {19322} (\bibinfo {year} {2013})}\BibitemShut {NoStop}%
\bibitem [{\citenamefont {Pedersen}\ \emph {et~al.}(2013)\citenamefont
  {Pedersen}, \citenamefont {Domesle}, \citenamefont {Lammich}, \citenamefont
  {Dziarzhytski}, \citenamefont {Guerassimova}, \citenamefont {Treusch},
  \citenamefont {Harbo}, \citenamefont {Heber}, \citenamefont {Jordon-Thaden},
  \citenamefont {Arion}, \citenamefont {Förstel}, \citenamefont {Stier},
  \citenamefont {Hergenhahn},\ and\ \citenamefont
  {Wolf}}]{pedersen_photolysis_2013}%
  \BibitemOpen
  \bibfield  {author} {\bibinfo {author} {\bibfnamefont {H.~B.}\ \bibnamefont
  {Pedersen}}, \bibinfo {author} {\bibfnamefont {C.}~\bibnamefont {Domesle}},
  \bibinfo {author} {\bibfnamefont {L.}~\bibnamefont {Lammich}}, \bibinfo
  {author} {\bibfnamefont {S.}~\bibnamefont {Dziarzhytski}}, \bibinfo {author}
  {\bibfnamefont {N.}~\bibnamefont {Guerassimova}}, \bibinfo {author}
  {\bibfnamefont {R.}~\bibnamefont {Treusch}}, \bibinfo {author} {\bibfnamefont
  {L.~S.}\ \bibnamefont {Harbo}}, \bibinfo {author} {\bibfnamefont
  {O.}~\bibnamefont {Heber}}, \bibinfo {author} {\bibfnamefont
  {B.}~\bibnamefont {Jordon-Thaden}}, \bibinfo {author} {\bibfnamefont
  {T.}~\bibnamefont {Arion}}, \bibinfo {author} {\bibfnamefont
  {M.}~\bibnamefont {Förstel}}, \bibinfo {author} {\bibfnamefont
  {M.}~\bibnamefont {Stier}}, \bibinfo {author} {\bibfnamefont
  {U.}~\bibnamefont {Hergenhahn}}, \ and\ \bibinfo {author} {\bibfnamefont
  {A.}~\bibnamefont {Wolf}},\ }\href {\doibase 10.1103/PhysRevA.87.013402}
  {\bibfield  {journal} {\bibinfo  {journal} {Physical Review A}\ }\textbf
  {\bibinfo {volume} {87}} (\bibinfo {year} {2013}),\
  10.1103/PhysRevA.87.013402}\BibitemShut {NoStop}%
\bibitem [{\citenamefont {Rajgara}\ \emph {et~al.}(2009)\citenamefont
  {Rajgara}, \citenamefont {Dharmadhikari}, \citenamefont {Mathur},\ and\
  \citenamefont {Safvan}}]{rajgara_strong_2009}%
  \BibitemOpen
  \bibfield  {author} {\bibinfo {author} {\bibfnamefont {F.~A.}\ \bibnamefont
  {Rajgara}}, \bibinfo {author} {\bibfnamefont {A.~K.}\ \bibnamefont
  {Dharmadhikari}}, \bibinfo {author} {\bibfnamefont {D.}~\bibnamefont
  {Mathur}}, \ and\ \bibinfo {author} {\bibfnamefont {C.~P.}\ \bibnamefont
  {Safvan}},\ }\href {\doibase 10.1063/1.3157234} {\bibfield  {journal}
  {\bibinfo  {journal} {The Journal of Chemical Physics}\ }\textbf {\bibinfo
  {volume} {130}},\ \bibinfo {pages} {231104} (\bibinfo {year}
  {2009})}\BibitemShut {NoStop}%
\bibitem [{\citenamefont {Streeter}\ \emph {et~al.}(2017)\citenamefont
  {Streeter}, \citenamefont {Yip}, \citenamefont {Reedy}, \citenamefont
  {Landers},\ and\ \citenamefont {McCurdy}}]{streeter_classical_2017}%
  \BibitemOpen
  \bibfield  {author} {\bibinfo {author} {\bibfnamefont {Z.}~\bibnamefont
  {Streeter}}, \bibinfo {author} {\bibfnamefont {F.}~\bibnamefont {Yip}},
  \bibinfo {author} {\bibfnamefont {D.}~\bibnamefont {Reedy}}, \bibinfo
  {author} {\bibfnamefont {A.}~\bibnamefont {Landers}}, \ and\ \bibinfo
  {author} {\bibfnamefont {W.}~\bibnamefont {McCurdy}},\ }\href@noop {}
  {\bibfield  {journal} {\bibinfo  {journal} {Bulletin of the American Physical
  Society}\ ,\ \bibinfo {pages} {D1.00135}} (\bibinfo {year}
  {2017})}\BibitemShut {NoStop}%
\bibitem [{\citenamefont {Richardson}\ \emph {et~al.}(1986)\citenamefont
  {Richardson}, \citenamefont {Eland}, \citenamefont {Fournier},\ and\
  \citenamefont {Cooper}}]{richardson_spectrum_1986}%
  \BibitemOpen
  \bibfield  {author} {\bibinfo {author} {\bibfnamefont {P.~J.}\ \bibnamefont
  {Richardson}}, \bibinfo {author} {\bibfnamefont {J.~H.~D.}\ \bibnamefont
  {Eland}}, \bibinfo {author} {\bibfnamefont {P.~G.}\ \bibnamefont {Fournier}},
  \ and\ \bibinfo {author} {\bibfnamefont {D.~L.}\ \bibnamefont {Cooper}},\
  }\href {\doibase 10.1063/1.450808} {\bibfield  {journal} {\bibinfo  {journal}
  {The Journal of Chemical Physics}\ }\textbf {\bibinfo {volume} {84}},\
  \bibinfo {pages} {3189} (\bibinfo {year} {1986})}\BibitemShut {NoStop}%
\bibitem [{\citenamefont {Gervais}\ \emph {et~al.}(2009)\citenamefont
  {Gervais}, \citenamefont {Giglio}, \citenamefont {Adoui}, \citenamefont
  {Cassimi}, \citenamefont {Duflot},\ and\ \citenamefont
  {Galassi}}]{gervais_h2o2+_2009}%
  \BibitemOpen
  \bibfield  {author} {\bibinfo {author} {\bibfnamefont {B.}~\bibnamefont
  {Gervais}}, \bibinfo {author} {\bibfnamefont {E.}~\bibnamefont {Giglio}},
  \bibinfo {author} {\bibfnamefont {L.}~\bibnamefont {Adoui}}, \bibinfo
  {author} {\bibfnamefont {A.}~\bibnamefont {Cassimi}}, \bibinfo {author}
  {\bibfnamefont {D.}~\bibnamefont {Duflot}}, \ and\ \bibinfo {author}
  {\bibfnamefont {M.~E.}\ \bibnamefont {Galassi}},\ }\href {\doibase
  10.1063/1.3157164} {\bibfield  {journal} {\bibinfo  {journal} {The Journal of
  Chemical Physics}\ }\textbf {\bibinfo {volume} {131}},\ \bibinfo {pages}
  {024302} (\bibinfo {year} {2009})}\BibitemShut {NoStop}%
\bibitem [{\citenamefont {Van~Huis}\ \emph {et~al.}(1999)\citenamefont
  {Van~Huis}, \citenamefont {Wesolowski}, \citenamefont {Yamaguchi},\ and\
  \citenamefont {Schaefer}}]{van_huis_scratching_1999}%
  \BibitemOpen
  \bibfield  {author} {\bibinfo {author} {\bibfnamefont {T.~J.}\ \bibnamefont
  {Van~Huis}}, \bibinfo {author} {\bibfnamefont {S.~S.}\ \bibnamefont
  {Wesolowski}}, \bibinfo {author} {\bibfnamefont {Y.}~\bibnamefont
  {Yamaguchi}}, \ and\ \bibinfo {author} {\bibfnamefont {H.~F.}\ \bibnamefont
  {Schaefer}},\ }\href {\doibase 10.1063/1.479127} {\bibfield  {journal}
  {\bibinfo  {journal} {The Journal of Chemical Physics}\ }\textbf {\bibinfo
  {volume} {110}},\ \bibinfo {pages} {11856} (\bibinfo {year}
  {1999})}\BibitemShut {NoStop}%
\bibitem [{\citenamefont {Nobusada}\ and\ \citenamefont
  {Tanaka}(2000)}]{nobusada_theoretical_2000}%
  \BibitemOpen
  \bibfield  {author} {\bibinfo {author} {\bibfnamefont {K.}~\bibnamefont
  {Nobusada}}\ and\ \bibinfo {author} {\bibfnamefont {K.}~\bibnamefont
  {Tanaka}},\ }\href {\doibase 10.1063/1.481374} {\bibfield  {journal}
  {\bibinfo  {journal} {The Journal of Chemical Physics}\ }\textbf {\bibinfo
  {volume} {112}},\ \bibinfo {pages} {7437} (\bibinfo {year}
  {2000})}\BibitemShut {NoStop}%
\bibitem [{\citenamefont {Winkoun}\ \emph {et~al.}(1988)\citenamefont
  {Winkoun}, \citenamefont {Dujardin}, \citenamefont {Hellner},\ and\
  \citenamefont {Besnard}}]{winkoun_one-_1988}%
  \BibitemOpen
  \bibfield  {author} {\bibinfo {author} {\bibfnamefont {D.}~\bibnamefont
  {Winkoun}}, \bibinfo {author} {\bibfnamefont {G.}~\bibnamefont {Dujardin}},
  \bibinfo {author} {\bibfnamefont {L.}~\bibnamefont {Hellner}}, \ and\
  \bibinfo {author} {\bibfnamefont {M.~J.}\ \bibnamefont {Besnard}},\ }\href
  {\doibase 10.1088/0953-4075/21/8/011} {\bibfield  {journal} {\bibinfo
  {journal} {Journal of Physics B: Atomic, Molecular and Optical Physics}\
  }\textbf {\bibinfo {volume} {21}},\ \bibinfo {pages} {1385} (\bibinfo {year}
  {1988})}\BibitemShut {NoStop}%
\bibitem [{\citenamefont {Schneider}, \citenamefont {Di~Giacomo},\ and\
  \citenamefont {Gianturco}(1996)}]{schneider_lower_1996}%
  \BibitemOpen
  \bibfield  {author} {\bibinfo {author} {\bibfnamefont {F.}~\bibnamefont
  {Schneider}}, \bibinfo {author} {\bibfnamefont {F.}~\bibnamefont
  {Di~Giacomo}}, \ and\ \bibinfo {author} {\bibfnamefont {F.~A.}\ \bibnamefont
  {Gianturco}},\ }\href {\doibase 10.1063/1.472582} {\bibfield  {journal}
  {\bibinfo  {journal} {The Journal of Chemical Physics}\ }\textbf {\bibinfo
  {volume} {105}},\ \bibinfo {pages} {7560} (\bibinfo {year}
  {1996})}\BibitemShut {NoStop}%
\bibitem [{\citenamefont {Brundle}\ and\ \citenamefont
  {Turner}(1968)}]{brundle_high_1968}%
  \BibitemOpen
  \bibfield  {author} {\bibinfo {author} {\bibfnamefont {C.~R.}\ \bibnamefont
  {Brundle}}\ and\ \bibinfo {author} {\bibfnamefont {D.~W.}\ \bibnamefont
  {Turner}},\ }\href {\doibase 10.1098/rspa.1968.0172} {\bibfield  {journal}
  {\bibinfo  {journal} {Proceedings of the Royal Society A: Mathematical,
  Physical and Engineering Sciences}\ }\textbf {\bibinfo {volume} {307}},\
  \bibinfo {pages} {27} (\bibinfo {year} {1968})}\BibitemShut {NoStop}%
\bibitem [{\citenamefont {Sanderson}\ \emph {et~al.}(1999)\citenamefont
  {Sanderson}, \citenamefont {El-Zein}, \citenamefont {Bryan}, \citenamefont
  {Newell}, \citenamefont {Langley},\ and\ \citenamefont
  {Taday}}]{sanderson_geometry_1999}%
  \BibitemOpen
  \bibfield  {author} {\bibinfo {author} {\bibfnamefont {J.~H.}\ \bibnamefont
  {Sanderson}}, \bibinfo {author} {\bibfnamefont {A.}~\bibnamefont {El-Zein}},
  \bibinfo {author} {\bibfnamefont {W.~A.}\ \bibnamefont {Bryan}}, \bibinfo
  {author} {\bibfnamefont {W.~R.}\ \bibnamefont {Newell}}, \bibinfo {author}
  {\bibfnamefont {A.~J.}\ \bibnamefont {Langley}}, \ and\ \bibinfo {author}
  {\bibfnamefont {P.~F.}\ \bibnamefont {Taday}},\ }\href {\doibase
  10.1103/PhysRevA.59.R2567} {\bibfield  {journal} {\bibinfo  {journal}
  {Physical Review A}\ }\textbf {\bibinfo {volume} {59}},\ \bibinfo {pages}
  {R2567} (\bibinfo {year} {1999})}\BibitemShut {NoStop}%
\bibitem [{\citenamefont {Légaré}\ \emph {et~al.}(2005)\citenamefont
  {Légaré}, \citenamefont {Lee}, \citenamefont {Litvinyuk}, \citenamefont
  {Dooley}, \citenamefont {Wesolowski}, \citenamefont {Bunker}, \citenamefont
  {Dombi}, \citenamefont {Krausz}, \citenamefont {Bandrauk}, \citenamefont
  {Villeneuve},\ and\ \citenamefont {Corkum}}]{legare_laser_2005}%
  \BibitemOpen
  \bibfield  {author} {\bibinfo {author} {\bibfnamefont {F.}~\bibnamefont
  {Légaré}}, \bibinfo {author} {\bibfnamefont {K.~F.}\ \bibnamefont {Lee}},
  \bibinfo {author} {\bibfnamefont {I.~V.}\ \bibnamefont {Litvinyuk}}, \bibinfo
  {author} {\bibfnamefont {P.~W.}\ \bibnamefont {Dooley}}, \bibinfo {author}
  {\bibfnamefont {S.~S.}\ \bibnamefont {Wesolowski}}, \bibinfo {author}
  {\bibfnamefont {P.~R.}\ \bibnamefont {Bunker}}, \bibinfo {author}
  {\bibfnamefont {P.}~\bibnamefont {Dombi}}, \bibinfo {author} {\bibfnamefont
  {F.}~\bibnamefont {Krausz}}, \bibinfo {author} {\bibfnamefont {A.~D.}\
  \bibnamefont {Bandrauk}}, \bibinfo {author} {\bibfnamefont {D.~M.}\
  \bibnamefont {Villeneuve}}, \ and\ \bibinfo {author} {\bibfnamefont {P.~B.}\
  \bibnamefont {Corkum}},\ }\href {\doibase 10.1103/PhysRevA.71.013415}
  {\bibfield  {journal} {\bibinfo  {journal} {Physical Review A}\ }\textbf
  {\bibinfo {volume} {71}} (\bibinfo {year} {2005}),\
  10.1103/PhysRevA.71.013415}\BibitemShut {NoStop}%
\bibitem [{\citenamefont {Cooper}\ \emph {et~al.}(1987)\citenamefont {Cooper},
  \citenamefont {Gerratt}, \citenamefont {Raimondi},\ and\ \citenamefont
  {Sironi}}]{cooper_study_1987}%
  \BibitemOpen
  \bibfield  {author} {\bibinfo {author} {\bibfnamefont {D.~L.}\ \bibnamefont
  {Cooper}}, \bibinfo {author} {\bibfnamefont {J.}~\bibnamefont {Gerratt}},
  \bibinfo {author} {\bibfnamefont {M.}~\bibnamefont {Raimondi}}, \ and\
  \bibinfo {author} {\bibfnamefont {M.}~\bibnamefont {Sironi}},\ }\href
  {\doibase 10.1063/1.453230} {\bibfield  {journal} {\bibinfo  {journal} {The
  Journal of Chemical Physics}\ }\textbf {\bibinfo {volume} {87}},\ \bibinfo
  {pages} {1666} (\bibinfo {year} {1987})}\BibitemShut {NoStop}%
\bibitem [{\citenamefont {Hishikawa}, \citenamefont {Takahashi},\ and\
  \citenamefont {Matsuda}(2006)}]{hishikawa_electronic_2006}%
  \BibitemOpen
  \bibfield  {author} {\bibinfo {author} {\bibfnamefont {A.}~\bibnamefont
  {Hishikawa}}, \bibinfo {author} {\bibfnamefont {E.~J.}\ \bibnamefont
  {Takahashi}}, \ and\ \bibinfo {author} {\bibfnamefont {A.}~\bibnamefont
  {Matsuda}},\ }\href {\doibase 10.1103/PhysRevLett.97.243002} {\bibfield
  {journal} {\bibinfo  {journal} {Physical Review Letters}\ }\textbf {\bibinfo
  {volume} {97}} (\bibinfo {year} {2006}),\
  10.1103/PhysRevLett.97.243002}\BibitemShut {NoStop}%
\end{thebibliography}%

\end{document}